\documentclass[11pt]{article}
\usepackage{graphicx}
\def\title{\begin{center}\Large\bf}
\def\author(s){\vspace{0.3cm}\large\rm}
\def\text{\end{center}}

\begin{document}


\noindent { \small{\it
\begin{center}
Contribution to the Conference "Interacting Binaries: Accretion and
Synchronization" \\ Crimean Astrophysical Observatory, June 20-26,
2008
\end{center}
}}
\hrule
\medskip



\title
Several results concerning the last stages of evolution of close
binaries composed of compact companions

\bigskip



\author(s)
B.E. Zhilyaev $^1$, D.L. Dubinovska $^2$

\bigskip

\smallskip

\noindent $^1${\small {\it Main Astronomical Observatory, NAS
of Ukraine,  27 Zabolotnoho, 03680 Kiev, Ukraine}} \\

\noindent {\small {\it e-mail:}} {\small{\bf zhilyaev@mao.kiev.ua}}

\smallskip

\smallskip

\noindent $^2${\small {\it Astrophysical Institute Potsdam, An der
Sternwarte 16, Potsdam 14482, Germany}}


\smallskip

\text

\noindent {\small {\bf keywords} gamma rays: bursts -- binaries:
close -- stars: neutron -- black hole physics -- methods:
statistical}

\section*{\small {Abstract}}

\small{\bf Gamma-ray bursts (GRB) are the most powerful transient
phenomena in the Universe. Nowadays dozens of speculations on the
origin of GRB were undertaken, but so far a single model for the
origin of, in particular, short GRBs does not exist. The black hole
(BH) - neutron star (NS) coalescence is a promising candidate source
for short GRBs. Most of binary mergers numerical simulations were
carried out with the purpose of investigating the emission of
gravitational waves. Such a scenario consists of an inspiral,
merging and ringdown phase. In this paper we present the comparison
of the observational results and analytical predictions for a test
particle in a quasicircular orbit around the BH. The emission of
gravitational waves causes a rapid decrease of the orbital radius
and a rise of a {\it chirp} of radiation. Matter orbiting the black
hole would be expected to produce high-frequency oscillations (HFO).
Timescales of the coalescence process are of the order of
milliseconds and oscillation frequencies of hundreds Hz for a system
with a solar mass BH companion. We report on the detection of HFO in
two short gamma-ray bursts in this paper. The frequencies and
durations of the oscillations are in agreement with the predicted
values. A {\it chirp} phenomenon is identified also. We therefore
argue in favor of BH-NS mergers as a scenario for the production of
short gamma-ray bursts.}

\bigskip

\section{Introduction}   

Gravitational wave (GW) astronomy is an inceptive branch of
observational astronomy. There exist many aims for observing with GW
detectors (LIGO, VIRGO, TAMA, GEO 600, e.t.c.), e.s. binary black
hole - black hole (BH-BH), neutron star - black hole (NS-BH), white
dwarf - black hole (WD-BH) systems. There are many evidences for the
presence of Supermassive Black Holes in the centers of many galactic
bulges. One is the detection of flares from tidally disrupted stars.
In the disruption process GW are also emitted. \noindent The
coalescence of the black hole (BH) with the neutron star (NS) can
probably also produce short GRBs with durations of milliseconds.

\noindent Most results for the binary mergers obtained from
relativistic numerical simulations were carried out with the purpose
of investigating the emission of gravitational waves. For such a
scenario the inspiral, merging and ringdown phases are typical.
\noindent As an example we can consider the evolution of a binary
system of two nonspinning black holes of equal masses $M_{0}$, with
an initial proper separation of approximately 16.6 $M_{0}$
(Pretorius 2005). Here we use c = G = 1. The binary is merged within
approximately 1 orbit, leaving behind a black hole mass of $ M
\simeq 1.9 \,M_{0}$. For $M_{0}= M_{\odot}$ the time/space scales
are equal to $250 \,\mu s$ and $148 \,km$, respectively.
Approximately 15\% of the total scalar field energy does not
collapse into black hole. The residual scalar field leaves the
vicinity of the orbit quite rapidly, within of the order of the
light crossing time of the orbit.  About 5 \% of the total mass will
be emitted as the gravitational waves during the final stages of the
collision (ringdown phase) lasting only a few milliseconds.

Faber et al. (2005) have performed relativistic calculations of
BH-NS mergers on the assumption that the BH is much more massive
than the NS. Adiabatic evolution calculations for neutron stars with
low compactness show that the neutron star typically disrupts
completely within a few orbital periods. The majority of the mass
transferred onto the black hole is accreted promptly; however a
significant fraction ($\simeq30$\%) of the mass moves outward, some
of which will become gravitationally unbound and ejected completely
from the system. The remaining portion forms an accretion disk
around the black hole, and could provide the energy source for
short-duration gamma-ray bursts. As matter accretes onto the BH, it
will excite quasinormal ringing modes that could in principle be
detected both as gravitational waves and electromagnetic radiation.
Gravitational wave emission calculations performed by Faber et al.
(2005) showed also the {\it "chirping"} signal, in which the binary
separation decreases while the GW amplitude and frequency increases.
This lasts until the onset of mass transfer from the neutron star
onto the black hole. At this point we are coming across a much more
rapid {\it "reverse chirp"}, as the gravitational wave amplitude and
frequency rapidly decrease while the NS is tidally disrupted.

\noindent In recent years SWIFT has detected the afterglows of the
short bursts GRB050724, GRB050813, and several others. These bursts
together provide evidence for the fact that some short GRBs are
occurring due to the merging of NS-NS binary systems (Gehrels \&
Leonard 2007).

In this paper we present the comparison of the numerical results and
analytical predictions based on the detailed calculations for a test
particle in a quasicircular orbit rotating about the BH. We report
on the detection of high-frequency oscillations during short
gamma-ray bursts. These phenomena might be evidence for to the
coalescence of a BH-NS binary.

\section{Background hypotheses}

A luminous gamma/X-ray burst can occur when a star passes within the
tidal radius of the massive black hole and is disrupted. Disruption
begins when the tidal acceleration by the black hole equals the
self-gravity of the star. It can be assumed that so far as the Roche
lobe lies within the star, mass loss lasts until a star is
completely disrupted. The tidal disruption timescale is about of the
free-falling time $T_{ff}$ . The characteristic time it will take a
body to collapse under its own gravity, if no other forces existed
to oppose the collapse is
\begin{equation}\label{free}
    T_{ff} = \frac{1}{4}\sqrt{\frac{3\pi}{2G\rho}}
\end{equation}
where $\rho$ is the mean density. Note also that the free-falling
timescale practically coincides with the oscillation period of a
self-gravitating body
\begin{equation}\label{osscil}
    P\approx \sqrt{\frac{4\pi}{G\rho}}
\end{equation}

Numerical calculations of the tidal disruption by Evans \& Kochanek
(1989) yielded disruption timescale close to above mentioned values
also. For the Sun $T_{ff}= 1.78\cdot 10^{3}$ s = 29.7 min; for a red
dwarf of M5 V class $T_{ff} = 685$ s = 11.8 min; for a white dwarf
and a neutron star of solar mass $T_{ff} = 1.78$ s and $\approx
0.0001 = 0.1 $ ms, respectively. Assuming this we can suggest that
short-duration gamma-ray bursts, with durations in the milliseconds
range, result from the tidal disruption of a neutron star by a black
hole.

Thus, a possible scenario for the production of HFO is connected to
the coalescence of stellar-mass black holes and neutron stars.
Matter orbiting a black hole produces periodic phenomena, which may
be identified with HFO. Such coalescence emits both gravitational
waves and gamma-rays due to the very hot gas that arises as a result
of the tidal disruption of a neutron star. The emission of
gravitational waves causes a rapid decrease of the orbital radius
and produces a \textit{chirp} of radiation (Shutz 2001). The details
of the GW waveforms (and luminosity) depend on geometrical
characteristics, masses and orientation of the binary. They also
show striking dependence on the stiffness of the equation of state
of the neutron matter which has been shown in numerical simulations
(Lee 2000, Rosswog 2004). Depending on the polytropic index the
radiation signal may reveal periodic peaks (HFO) or drop abruptly to
zero after the star is disrupted.

It appears that HFO contain encoded data about both the geometry and
physics of the BH-NS binary. Of interest is an important feature of
a Reissner-Nordstr\"{o}m black hole, namely the possibility to
convert the reflected gravitational waves arising from the
coalescence process to electromagnetic radiation (Chandrasekhar
1983). The predicted rate from such a conversion may amount to a few
tens of percents at resonant frequencies that depend on the BH size
(Gunter 1981).

The gravitational-wave luminosity of two stars of equal mass $M$ in
an orbit of radius $R$ is in order of magnitude (Shutz 2001)
\begin{equation}\label{L}
   L \simeq \frac{1}{80}\frac{c^{5}}{G}\left( \frac{GM}{Rc^{2}} \right)^{5}
\end{equation}
It is possible to calculate that close BH-NS binaries with $M =
M_{\odot}$ and $R < 1/6 \,R_{\odot}$ can radiate more energy in
gravitational waves than the Sun in light. Then note that the
gravitational wave amplitude for a binary system is
\begin{equation}\label{hb}
   h_{binary} \simeq \frac{1}{2} \frac{GM}{r\,c^{2}}\frac{GM}{R\,c^{2}}= \frac{1}{8}\frac{r_{g}^{2}}{r\,R}
\end{equation}
Perturbations of the metric in the vicinity of coalescing stars $(r
\sim R \sim r_{g}$, where $r_{g} = 2GM/c^{2}$ is the gravitational
radius), may be extremely large $(h \simeq 1)$. An important point
is that the variations of gamma-ray intensity will carry a print of
strong gravitational fields from in-spiraling orbits (gravitational
waveforms) on them. It may provide another insight into mergers of
stars and BHs. These events are expected to be detected by future
space-based gravitational-wave interferometers. Remarkably, similar
results can be obtained, analyzing gamma-ray HFO.

Thus, the most simple model suggests that the frequency of HFO would
be equal to the local Keplerian frequency $f =
(GM/R^{3})^{1/2}/2\pi$. Detailed calculations (Chandrasekhar 1983;
Novikov \& Frolov 1986)have shown, that for a test particle of mass
$m$ the stable circular orbits around the BH of mass $M$ exist for
$R > 3\, r_{g}$ only. The binding energy of the test particle in
close proximity to the critical orbit is $E \simeq 0.06 \,mc^{2}$.
During each cycle in the critical orbit the particle radiates the
gravitational energy $\Delta E \approx 0.1\, mc^{2}(m/M)$. In
general, inside the critical orbit the particle moves in a spiral
curve with $N$ cycles before coalescence, where $N \approx
(M/m)^{1/3}$. Gravitational wave radiation causes the orbital radius
to shrink, thus for $R \gg r_{g}$ it has the following form (Landau
\& Lifshitz 1973)
\begin{equation}\label{dRdt}
    \frac{dR}{dt} =- \frac{8}{5}\,c \left ( \frac{m}{M}\right)\left (\frac{r_{g}}{R}
    \right)^{3}
\end{equation}
If the initial radius of the Newtonian orbit $R$ of two stars of the
equal mass $M_{\odot}$ ranges from 4 to 10 gravitational radii, i.e.
approximately 12 to 30 km, the frequency of HFO will lay between 350
Hz and 2.2 kHz. In this case a close binary appears to be extremely
unstable. The gravitational-wave energy released around the critical
orbit can be as high as $10^{53}$ ergs. Integrating the differential
equation (5) in time we will get the survival time of a binary
system with initial conditions mentioned above. This time is around
0.015 s. Strikingly, the gamma-ray bursts considered here
demonstrate both the HFO frequencies and the times of oscillations
close to the mentioned values.

Note that all the above estimates should be treated as approximate
results. These are valid for test particles orbiting black holes.
Coalescence of the BH-NS binaries requires the solution of the set
of Einstein equations. To obtain the right expression for the
orbital frequency, a calculation in general relativity is necessary.
The case of a circular orbit $R = 3\, r_{g}$ corresponds to a
singular trajectory. It has no analogue in Newtonian theory.
Detailed relativistic calculations show that the local Keplerian
period $T_{K}$ and the rotational period $T$ for an outside observer
are related as (Chandrasekhar 1983)
\begin{equation}\label{Tk}
T_{K}= \left ( 4\pi ^{2} R^{3}/ GM \right )^{1/2}
\end{equation}
\begin{equation}\label{T}
    T = T_{K}(1-6\,\mu)^{-1/2} = \pi \frac{r_{g}}{c}[\mu^{3}(1-6\,
\mu)]^{-1/2}
\end{equation}
%
where $\mu = r_{g} / 2R$. It can be shown, that while an orbital
radius $R$ is larger than $4\,r_{g}$ the observed frequency will be
{\it chirping up} (see Fig. 1). For $R < 4\,r_{g}$ it goes down
({\it "reverse chirp"}). This can be explained by the time dilation
close to the event horizon of the black hole. The observed period
$T$ diverges to infinity when $R$ approaches $3\, r_{g}$. Thus, the
chirping timescale determines both the binary mass and size.
\begin{figure}[!ht]
\resizebox{1.00\hsize}{!}{\includegraphics{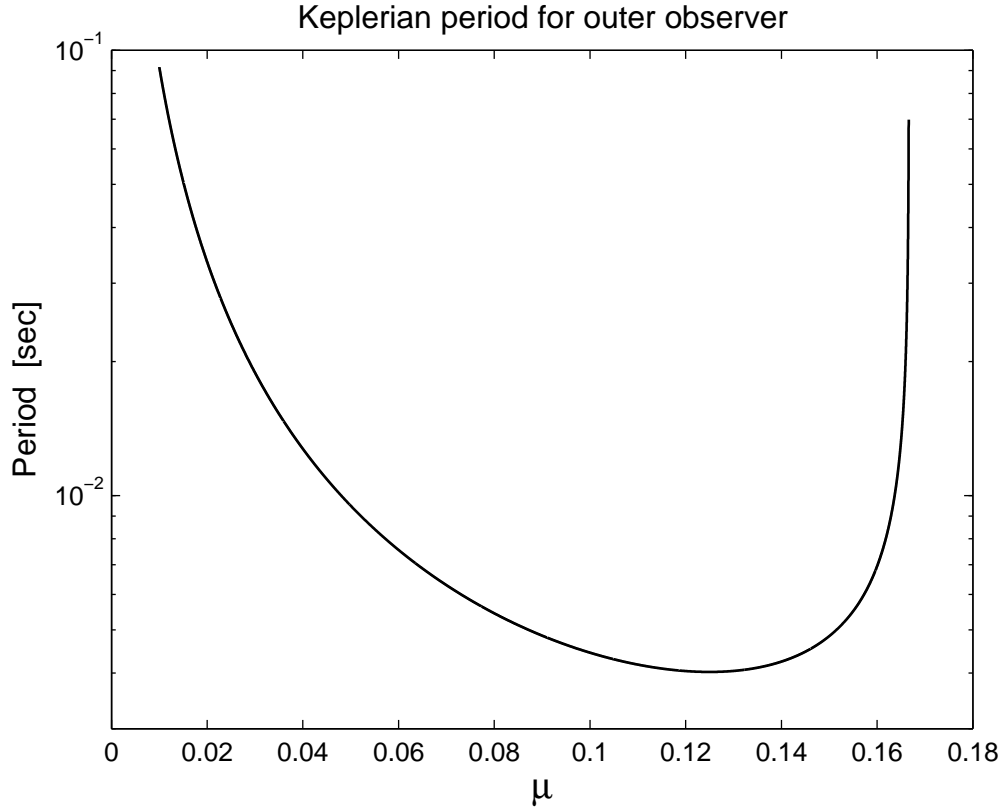}}
\caption{{\small The variation of the Keplerian rotational period
for an external observer as derived from equation (7). The abscissa
units are
 $\mu = r_{g} / 2R$ for $M=M_{\odot}$. The scale of the ordinate is
given in seconds. }}
\end{figure}
\begin{figure}[!ht]
\resizebox{1.00\hsize}{!}{\includegraphics{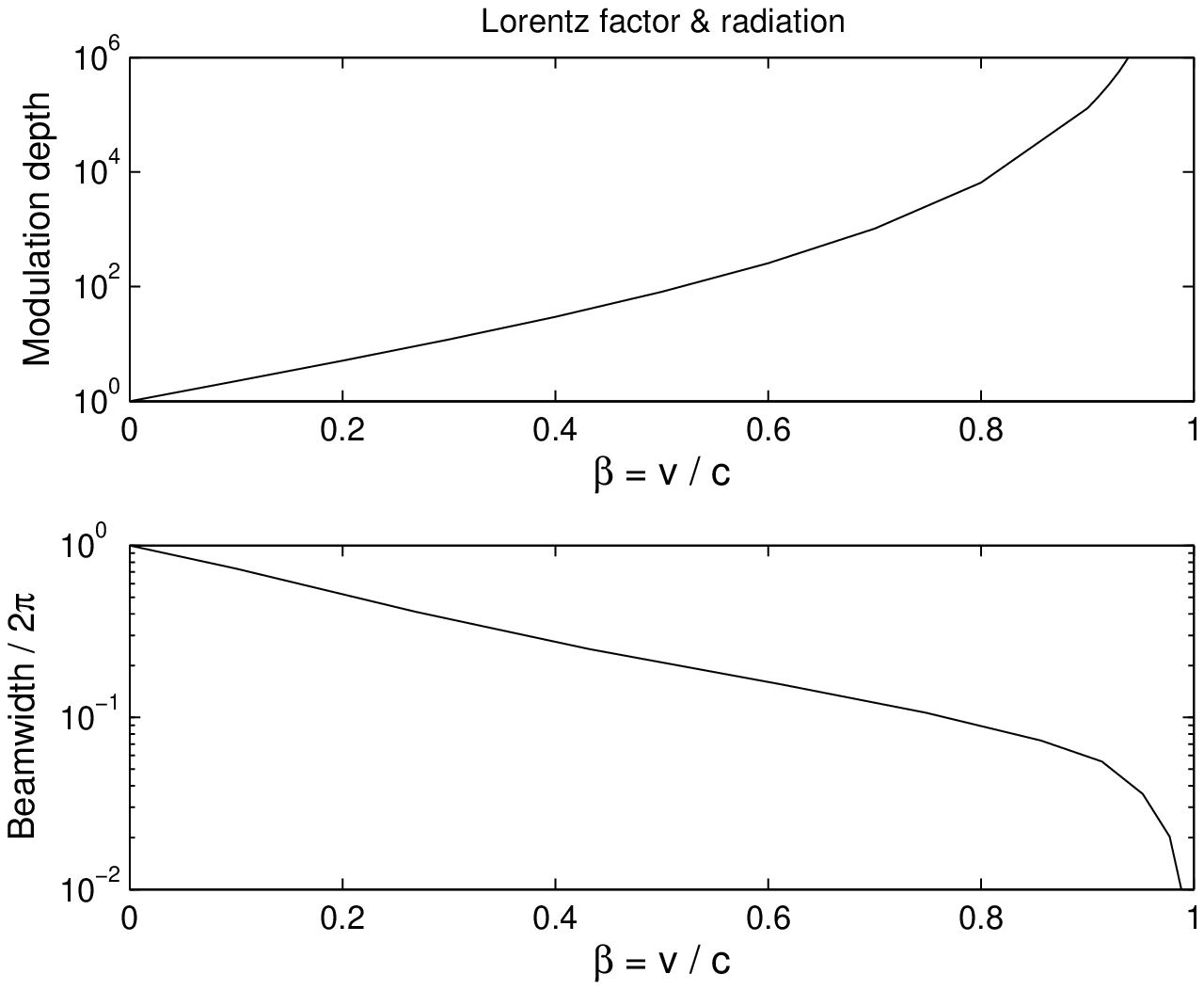}}%
\caption{{\small Variations in both modulation depth and half-power
beamwidth of the HFO signal as function of $\beta$ as follows from
equation (7).}}
\end{figure}

Lee \& Ramirez-Ruiz (2002) studied numerically the hydrodynamic
evolution of massive accretion disks formed as a result of the
disruption of a neutron star by a black hole in the context of
gamma-ray bursts formation. A gas mass $\simeq 0.1-0.25\, M_{\odot}$
survives in the orbiting debris, which enables strong magnetic
fields $\simeq 10^{16}$ G to be anchored in the dense matter long
enough to power short duration GRBs. They estimate the energy
released from the system through magnetic-dominated mechanisms and
find it can be as high as $10^{51}$ ergs during an estimated
accretion timescale of 0.1...0.2 s.

Impact of two stars may drive their own oscillations. The amplitudes
of the forced oscillations may be large enough to induce a variety
of nonlinear effects, in particular, generation of both higher and
fractional harmonics, etc.

HFO may also be related to the fundamental pulsation frequency of
coalescing stars. The oscillation frequency of a self-gravitating
body is  $f\sim (G\bar{\rho}/4\pi)^{1/2} $ (Shutz 2001). This
quantity is of the same order as the Keplerian orbital frequency.
For BH-NS binaries with $M = M_{\odot}$ and $R_{NS}\simeq $ 10 km,
$R_{BH}= r_{g} = 2GM_{\odot}/c^{2} $= 2.95 km, the fundamental
pulsation frequency is 1.6 and 9.9 kHz, respectively. In such a case
HFO can be regarded as a probe for the structure and the equation of
state of neutron stars as well as an asteroseismology probe for
black holes.

The radiation from a relativistic object is a nontrivial problem,
which requires treating it relativistically using Lorentz transform.
The Lorentz factor $\gamma$ appears in Special Relativity. It is
defined as:
\begin{equation}\label{}
   \gamma \equiv \frac{1}{\sqrt{1-\beta^{2}}}\,;\,\, \beta =\frac{v}{c}
\end{equation}
where $c$ is the speed of light and  $v$ is the velocity of the
object. An object moving with respect to an observer will seem to
radiate in a different way because of the time dilation and length
contraction. Intensity for external observer is (Lightman et al.
1975)
\begin{equation}\label{}
    L_{observ}=\frac{L}{4\pi R^{2}}\,\frac{1}{\gamma^{4}(1+\frac{v}{c}\,cos(\theta))^{4}}
\end{equation}
where $L$ is the luminosity and $\theta$ is the angle between the
line of sight from object to observer and the direction of motion.
We can see that intensity $L_{observ}$ depends both on the Lorentz
factor $\gamma$ and the angle $\theta$. When $\beta$ is much less
than unity, the radiation pattern is symmetric both in the forward
and backward directions. However, as $\beta = v/c \rightarrow 1$,
the radiation pattern becomes more and more concentrated to the
forward direction. For a highly relativistic object, the radiation
is emitted in a narrow cone which axis is aligned to the direction
of motion.

HFO in GRBs may be treated as a result of intensity modulation
caused by the variation in an angle $\theta$ at a relativistic rates
of movement in a close binary system. Modulation depth may be
defined as the peak value of intensity divided by its minimum value.
We define also a beamwidth of varying intensity $L_{observ}$ as
$FWHM$ (full width at half maximum). Both the modulation depth and
beamwidth of the HFO signal are shown in Fig. 2 as function of
$\beta$ following equation (7).

So if $\beta = 0.5$, the HFO intensity changes by approximately two
orders of magnitude, and the half-power beamwidth  is equal to about
$30^{\circ}$. In that case we can observe the HFO phenomenon if the
inclination angle to the line of sight of a emitting system is more
than $60^{\circ}$.

Not all short bursts have identical features. Since at relativistic
velocities the radiation is emitted in a narrow cone, some bursts
will not show the HFO effect because of geometrical reasons.

\section{Individual GRB events: BATSE triggers \# 432 and 512}
We have performed searches for high-frequency periodicities in two
short gamma-ray bursts on millisecond timescales. In our analysis we
used the TTE data from the BATSE 3B catalogue (Meegan et. al. 1996).
The TTE data contains the arrival time over a 2 microseconds time
bin, the energy of each photon and the number of the detector.

For trigger number 423, Cline et al. (1999) give the $T_{90}$ burst
duration followed from TTE fit equal to $0.050 \pm 0.002$ s. They
also note the detailed structure of BATSE trigger 512, which has the
finest time structure of any GRB observed to date - possibly down to
the 20 $\mu s$ level. This is a very bright burst with maximum power
in the third energy channel (100-300 keV). Its duration is $0.014
\pm 0.0006$ s only.

We investigate the intensity oscillations in short gamma-ray bursts
with wavelet analysis (Torrence \& Compo 1998).

\begin{figure}[!ht]
\resizebox{1.00\hsize}{!}{\includegraphics{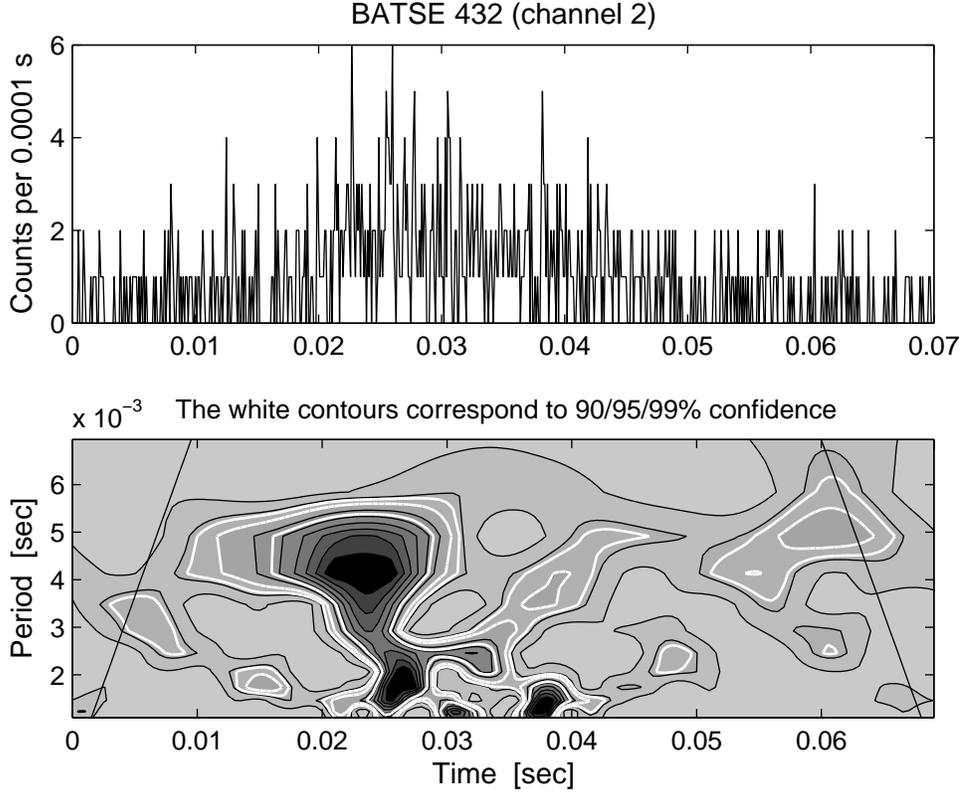}}%
\caption{{\small The light curve of BATSE trigger 432 in the energy
channel 50-100 keV from the TTE data binned to 100 $\mu s$
resolution (top) and its wavelet power spectrum (bottom). }}
\label{figure:BATSE432}
\end{figure}
\begin{figure}[!ht]
\begin{center}
\resizebox{0.90\hsize}{!}{\includegraphics{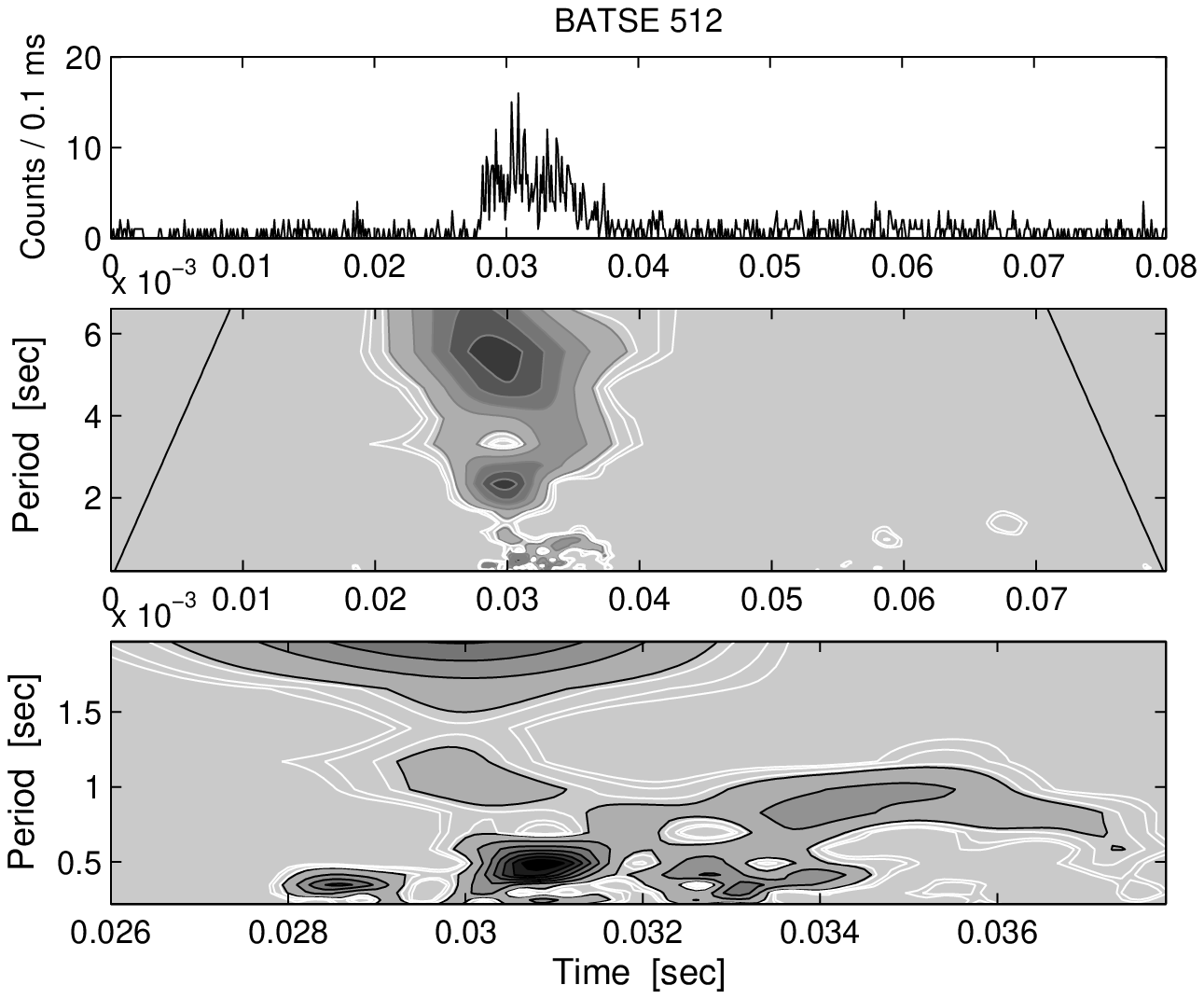}}%
\caption{{\small BATSE trigger 512. The upper panel shows the light
curve in the energy channel 100-300 keV with a sampling time of 100
$\mu s$. The middle panel shows the local wavelet power spectrum.
The lower panel with a magnified portion of the high-frequency
spectrum demonstrates a \textit{chirp} phenomenon. The outside white
contours enclose regions of power of greater than 95\% confidence
relative to white noise.}} \label{figure:BATSE512}
\end{center}
\end{figure}

Fig. 3 shows the light curve of BATSE trigger 432 in the energy
channel 50-100 keV and its wavelet power spectrum. The white
contours give the significance levels of 90\%, 95\% and 99\%,
respectively. The dark areas point to the burst oscillations with a
significance level of $>$ 99.9\%. This  spectrum reveals a
significant peak around 5 ms.

The plot shows also the {\it chirping} signal with an oscillation
period varying in time. The oscillation was detected at the burst
onset at a frequency of about 300 Hz, and it increased in frequency
over the following 30 ms of the burst rising to a maximum of about
1000 Hz. From that moment and forth the observed frequency is {\it
chirping} down. It is decreasing over the following 20 ms of the
burst decay to a minimum of about 400 Hz. As was mentioned above,
its decrease may be explained by the time dilation close to the
event horizon of a black hole. The bead-like structure of the
spectrum depends, in particular, on the unstable behavior of the
high-frequency components. This {\it chirping} phenomenon points to
a coalescing binary with a black hole companion of stellar mass.

\newpage

The wavelet transform allows us to extract the {\it chirping} signal
from the instrumental noise. Equation (5) gives the orbital
shrinking time of 0.015 s for the BH of the solar mass. This is in
agreement with an estimate of the burst duration from a fit to the
TTE light curve. The minimum period of the orbit in the range
$(1.5\ldots 3)\,10^{-3}$ s provides also the mass of BH  $\simeq
M_{\odot}$ according to equation (7). As discussed above in close
proximity to the critical orbit we can probably observe only a few
cycles of radiation before coalescence. For an orbital radius $R <
4\,r_{g}$ the observed frequency was found to be {\it chirping}
down.

Fig. 4 shows at least two curves {\it chirping} down in the local
wavelet power spectrum of BATSE trigger 512. As noted by Lee \&
Ramirez-Ruiz (2002), a gas mass $0.1\ldots 0.25\, M_{\odot}$
survives in the orbiting debris after the neutron star is disrupted
by the black hole. Multiple chirping curves might point to fragments
of different mass falling into th BH. Note that equation (6) for
local Keplerian period $T_{k}$ is valid only for the infinitesimal
mass particles orbiting black holes.

Fig. 5 also shows the decomposed light curves of BATSE trigger 512,
following the approach offered by Torrence \& Compo (1998). The thin
curve represents the high-frequency range, 0.2 ms $<$ period $<$ 1
ms. The thick curve encompasses the low-frequency diapason with
periods $>$ 1 ms. When added, they would form the original light
curve. The dashed lines mark the $\pm$ 3 sigma error corridor. Our
results referring to reconstruction can be summarized as follows.
\begin{itemize}
  \item The oscillations occur only during the burst phase. They appear
  suddenly with a rise time comparable with resolving time of 100
  microseconds. Recall that the dynamical time of the disruption of a NS
  as mentioned above is about 0.1 ms.
  \item Of interest for theory are oscillations that are visible in
  the both low- and high-frequency light curves. The HFO amplitude
  can be large, a significant fraction of the energy released is radiated in HFO.
  The contribution of HFO to the total luminosity of the burst event is
  estimated to be roughly up to 50\%.
  \newpage
\begin{figure}[!ht]
\begin{center}
\resizebox{.80\hsize}{!}{\includegraphics{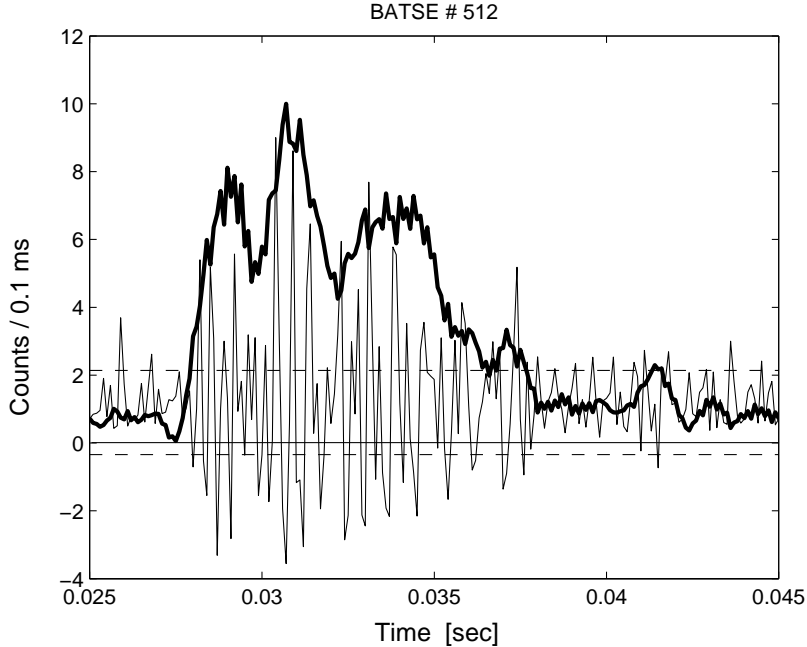}}%
\caption{{\small The decomposed light curve of BATSE trigger 512
(see text). The data of the energy channel 100-300 keV is used with
a resolving time of 100 $\mu s$ in two frequency bands. The thin
curve represents the high-frequency range (0.2 ms $<$ period $<$ 1
ms) with the $\pm$ 3 sigma error corridor (the dashed lines). The
thick curve encompasses the low-frequency diapason with periods $>$
1 ms.}  } \label{figure:BATSE512} \end{center}
\end{figure}
  \item The instantaneous frequency of HFO varies with time, as may
  be proved from the peak-to-peak time measurements in Fig. 5. It appears that
  the HFO amplitude changes periodically. The amplitude of modulation
  of oscillations amounts up to 100\%. As one might expect
  from equation (9) and Fig. 2 a radiating object moves with a velocity
  $\geq \, 0.6\,c$.
  \item The low-frequency light curve shows a few cycles with a mean
period between 1.9 ms and 4.4 ms, chirping down. The amplitude of
modulation  of low-frequency oscillations amounts  only to 30\%.
From here a radiating object moves with the velocity $\simeq \,
0.2\,c$.
\item It seems, we deal with fragments of matter moving
in orbits of different hight.
\end{itemize}

\section{Conclusion}   

The detection of high-frequency oscillations in short gamma-ray
bursts provides a new prospect for exploring the nature of this
mysterious phenomenon. HFO with periods in the range of milliseconds
and magnitudes of a few tens of percent of the burst luminosity may
be related to the accretion of debris formed after the tidal
disruption of a neutron star by a black hole in a coalescing binary
system. For such a scenario, one would expect a few cycles of
radiation at an orbital frequency {\it chirping up and down}. This
phenomenon indeed is seen in the bursts considered here.

\newpage



\begin{thebibliography}{}

\small{ \baselineskip=1pt

\bibitem{Cline}
{\it Cline, D.B., Matthey, C., \& Otwinowski, S.} , 1999,
arXiv:astro-ph/9905346, v1, 26 May 1999

\bibitem{Chandra}
{\it Chandrasekhar, S.} 1983, "The Mathematical Theory of Black
Holes", V. 1-2,(Oxford University Press)

\bibitem{Evans}
{\it Evans, C.R., \&  Kochanek, C.S.} 1989, ApJ, 346, L13-L16

\bibitem{Gehrels}
{\it Gehrels, N., \& Leonard, P.} 2007, EAS Newsletter, 33, 9

\bibitem{Gunter}
{\it Gunter, D.} 1981, Phil. Trans. Roy. Soc., 301, 705

\bibitem{Landau}
{\it Landau, L.D., Lifshitz, E.M.} 1973, "The Theory of Field",
(Moscow: Nauka)

\bibitem{Lee}
{\it Lee, W.H.} 2000, MNRAS, 318, 606

\bibitem{Lee02}
{\it Lee, W.H., \& Ramirez-Ruiz, E.} 2002, ApJ, 577, N 2, 893

\bibitem{Lightman}
{\it Lightman,A.P., Press,W.H., Price,R.H., \& Teukolsky, S.A. }
1975, "Problem Book in Relativity and Gravitation", (Princeton
University Press: Princeton, N.J.)

\bibitem{Meegan}
{\it Meegan, C. A., Pendleton, G.N., et al.} 1996, "The Third BATSE
Gamma-Ray Burst Catalog", ApJ Suppl., 106, 65

\bibitem{Novikov}
{\it Novikov, I.D., \& Frolov, V.P.} 1986, "Physics of Black Holes",
(Moscow: Nauka)

\bibitem{Pretorius}
{\it Pretorius, F.} 2005, arXiv:gr-qc/0507014 v1 4 Jul 2005

\bibitem{Rosswog}
{\it  Rosswog, S.,  Speith, R., \&  Wynn, G. A.} 2004, Mon. Not. R.
Astron. Soc. 351, 1121

\bibitem{Shutz}
{\it Shutz, B.} 2001, "Gravitational Radiation", Encyclopedia of
Astronomy and Astrophysics, (Nature Publishing Group 2001)

\bibitem{Torrence}
{\it Torrence, C., \& Compo, G. P.} 1998, Bull. Amer. Meteor. Soc.,
79, 6

}

\end{thebibliography}
\end{document}